\documentclass[twocolumn,           
               showpacs,            
               nopreprintnumbers,     
               aps,                 
               prd,          	    
               letterpaper,             
               groupeaddress,      
               nofootinbib,         
               tightenlines,        
               floats,floatfix      
               ]{revtex4-1}

\usepackage{graphicx}
\usepackage{dcolumn}
\usepackage{bm}
\usepackage{amsmath}
\usepackage{amsfonts,amssymb}
\usepackage{color}
\usepackage{enumerate}

\begin{document}

\title{Branes constrictions with White Dwarfs}

\author{Miguel A. Garc\'{\i}a-Aspeitia}
\email{aspeitia@fisica.uaz.edu.mx}

\affiliation{Consejo Nacional de Ciencia y Tecnolog\'ia, Av, Insurgentes Sur 1582. Colonia Cr\'edito Constructor, Del. Benito Ju\'arez C.P. 03940, M\'exico D.F. M\'exico}
\affiliation{Unidad Acad\'emica de F\'isica, Universidad Aut\'onoma de Zacatecas, Calzada Solidaridad esquina con Paseo a la Bufa S/N C.P. 98060, Zacatecas, M\'exico.}

\begin{abstract}
We consider here a robust study of stellar dynamics for White Dwarf Stars with polytropic matter in the weak field approximation using the Lane-Emden equation from the brane-world scenario. We also derive an analytical solution to the nonlocal energy density and show the behavior and sensitivity of these stars to the presence of extra dimensions. Similarly, we analyze its stability and compactness, in order to show whether it is possible to be close to the conventional wisdom of white dwarfs dynamics. Our results predicts an average value of brane tension as: $\langle\lambda\rangle\gtrsim84.818\;\rm MeV^4$, with a standard deviation $\sigma\simeq82.021\;\rm MeV^4$ which comes from a sample of dwarf stars, being weaker than other astrophysical observations but remaining above of cosmological results provided by nucleosynthesis among others.
\end{abstract}

\draft
\date{\today}
\maketitle

\section{Introduction} \label{Int}

Stellar astrophysics has been a cornerstone to demonstrate the predictive capabilities of the General Theory of Relativity (GR), describing high energy astrophysical phenomena such as white dwarfs and neutron stars with unprecedented success\cite{Chandrasekhar:1985kt,*tolman1987relativity,*oppenheimer1939massive}. One of the most important results in this vein, is the Lane-Emden (LE) equation\cite{chandrasekhar1958introduction,*weinberg1972gravitation}, which is a Newtonian approach to GR, under the assumption that the dwarf star is formed by polytropic matter; remarking, that these types of stars are excellent high energy laboratories with which it is possible to test the phenomena described by GR and even to corroborate or refute our most plausible extensions\cite{gm,*Ovalle:2013vna,*Ovalle:2013xla,*Ovalle:2014uwa,*Ovalle:2014uwa,*Garcia-Aspeitia:2014jda,*Garcia-Aspeitia:2014pna,*Linares:2015fsa}.

Moreover, brane-world theory (for a good review see\cite{PerezLorenzana:2005iv,mk}) has been one of the most captivating extensions to GR, due to its theoretical predictions and its ability to solve fundamental phenomena such as the hierarchy problem, among others\cite{Randall-I,Randall-II,PhysRevLett.104.141601}. It is worth mentioning that the brane-world models has a very long tradition in the specialized literature and their properties have been extensively studied under diverse circumstances, from the cosmological scenarios\cite{Maartens:2000fg,*Brax:2004xh,*Brax:2003fv,*Aspeitia:2009bj,*GarciaAspeitia:2011xv} to the study of astrophysical models\cite{gm}.

Following the conventional wisdom, it is possible to extend the classical astrophysics for polytropic stellar systems with the brane-worlds frame work. In this vein, many authors have been given the task of showing the different stellar behavior, studying stability, collapse\cite{gm,Garcia-Aspeitia:2014pna,*Linares:2015fsa}or the stellar dynamic in general\cite{Ovalle:2013vna,*Ovalle:2013xla,*Ovalle:2014uwa,*Castro:2014xza,*Casadio:2012pu,*Casadio:2012rf,*Ovalle:2014uwa}.

Under this scenario, this paper is devoted to study the modifications of LE equation caused by brane in the cases of stars with polytropic matter; being our main goal, produce observational verifications in these systems. It is important to remark that one of the most suitable signatures is the sensitivity of these kind of stars to the corrections provided by brane theory, producing a new dynamics in energy density (or pressure) and in the effective mass; as well as the implementation of a new range of exclusion, where the star is dynamically unstable. From this new range, it is possible to propose a bound to the brane tension in order to avoid an unstable stellar configuration among other pathologies.

Before starting, we would like to mention here some experimental constraints on brane-world models, most of them about the so-called brane tension $\lambda$, which appears explicitly as a free parameter in the corrections of the gravitational equations mentioned above. As a first example we have the measurements on the deviations from Newton's law of the gravitational interaction at small distances. It is reported that no deviation is observed for distances $l \gtrsim  0.1 \, {\rm mm}$, which then implies a lower limit on the brane tension in the model Randall-Sundrum II (RSII): $\lambda> 1 \, {\rm TeV}^{4}$\cite{Kapner:2006si,*Alexeyev:2015gja}; it is important to  mention that these limits do not apply to the two-branes case of the model Randall-Sundrum I (RSI) (see\cite{mk} for details). Astrophysical studies related with gravitational waves and stellar stability, constraint
  the brane tension as $\lambda > 5\times10^{8} \, {\rm MeV}^{4}$\cite{gm,Sakstein:2014nfa,*GA2013}, whereas the existence of black hole X-ray binaries suggest that $l\lesssim 10^{-2} {\rm mm}$\cite{mk,Kudoh:2003xz,*Cavaglia:2002si}. Finally, from cosmological observations, the requirement of successful nucleosynthesis provides the lower limit $\lambda> 1\, {\rm MeV}^{4}$, which is a much weaker limit as compared to other experiments (other cosmological tests can be seen in\cite{Holanda:2013doa,*Barrow:2001pi,*Brax:2003fv}).

We divide this paper in the following sections: Section \ref{EM} is dedicated to show the equations of motion for a stellar structure, showing the modified Tolman-Oppenheimer-Volkoff (TOV) equation and the respective conservation equations; considering always the regularity of the functions and maintaining a Schwarzschild stellar exterior\cite{Garcia-Aspeitia:2014pna,Linares:2015fsa}. In section \ref{LES} we derive the LE and mass equations, based on a set of minimal assumptions which are in concordance with the current studies of stellar dynamics. Also, it is derived an analytical form of the nonlocal energy density which essentially is a function of the polytropic constant and the interior central energy density of the star. In Sec. \ref{NA} are imposed the initial conditions and we generate numerical solutions to the LE and mass equations for the case with polytropic index $n=3$, related with white dwarf stars. Finally in Sec. \ref{C} we give some conclusions and important remarks.

Henceforth we will use units in which $\hbar=c=1$, unless it is explicitly mentioned.

\section{Equations of motion} \label{EM}

Let us start by writing the equations of motion for stellar stability in a brane embedded in a five-dimensional bulk according to the RSII model\cite{Randall-II}. Following an appropriate computation (for details see\cite{mk,sms}), it is possible to demonstrate that the modified four-dimensional Einstein's equations can be written as 
\begin{equation}
  G_{\mu\nu} + \xi_{\mu\nu} + \Lambda_{(4)}g_{\mu\nu} = \kappa^{2}_{(4)} T_{\mu\nu} + \kappa^{4}_{(5)} \Pi_{\mu\nu} +
  \kappa^{2}_{(5)} F_{\mu\nu} , \label{Eins}
\end{equation}
where $\kappa_{(4)}$ and $\kappa_{(5)}$ are respectively the four and five- dimensional coupling constants, which are related in the form: $\kappa^{2}_{(4)}=8\pi G_{N}=\kappa^{4}_{(5)} \lambda/6$, where $\lambda$ is defined as the brane tension, and $G_{N}$ is Newton constant. For purposes of simplicity, we will not consider bulk matter, which translates into $F_{\mu\nu}=0$, and discard the presence of the four-dimensional cosmological constant, $\Lambda_{(4)}=0$, as we do not expect it to have any important effect at astrophysical scales (for a recent discussion about it see\cite{Pavlidou:2013zha}). Additionally, we will neglect any nonlocal energy flux, which is allowed by the static spherically symmetric solutions we will study below\cite{gm}.

The energy-momentum tensor, the quadratic energy-momentum tensor, and the Weyl (traceless) contribution, have the explicit forms
\begin{subequations}
\label{eq:4}
\begin{eqnarray}
\label{Tmunu}
T_{\mu\nu} &=& \rho u_{\mu}u_{\nu} + p h_{\mu\nu} \, , \\
\label{Pimunu}
\Pi_{\mu\nu} &=& \frac{1}{12} \rho \left[ \rho u_{\mu}u_{\nu} + (\rho+2p) h_{\mu\nu} \right] \, , \\
\label{ximunu}
\xi_{\mu\nu} &=& - \frac{\kappa^4_{(5)}}{\kappa^4_{(4)}} \left[ \mathcal{U} u_{\mu}u_{\nu} + \mathcal{P}r_{\mu}r_{\nu}+ \frac{ h_{\mu\nu} }{3} (\mathcal{U}-\mathcal{P} ) \right] \, .
\end{eqnarray}
\end{subequations}
Here, $p$ and $\rho$ are, respectively, the pressure and energy density of the stellar matter of interest, $\mathcal{U}$ is the nonlocal energy density, and $\mathcal{P}$ is the nonlocal anisotropic stress. Also, $u_{\alpha}$ is the four-velocity (that also satisfies the condition $g_{\mu\nu}u^{\mu}u^{\nu}=-1$), $r_{\mu}$ is an unit radial vector, and $h_{\mu\nu} = g_{\mu\nu} + u_{\mu} u_{\nu}$ is the projection operator orthogonal to $u_{\mu}$.

Spherical symmetry indicates that the metric can be written as:
\begin{equation}
{ds}^{2}= - B(r){dt}^{2} + A(r){dr}^{2} + r^{2} (d\theta^{2} + \sin^{2} \theta d\varphi^{2}) \, .\label{metric}
\end{equation}
If we define the reduced Weyl functions $\mathcal{V} = 6 \mathcal{U}/\kappa^4_{(4)}$, and $\mathcal{N} = 4 \mathcal{P}/\kappa^4_{(4)}$, then the equations of motion for a relativistic star in the brane are:
\begin{subequations}
  \label{eq:7}
\begin{eqnarray}
  \mathcal{M}^\prime &=& 4\pi{r}^{2}\rho_{eff} \, \label{eq:7d} \, , \\
  p^\prime &=& -\frac{G_N}{r^{2}} \frac{4 \pi \, p_{eff} \, r^3 + \mathcal{M}}{1 - 2G_N \mathcal{M}/r} ( p + \rho ) \, ,   \label{eq:7b} \\
  \mathcal{V}^{\prime} + 3 \mathcal{N}^{\prime} &=& - \frac{2G_N}{r^{2}} \frac{4 \pi \, p_{eff} \, r^3 + \mathcal{M}}{1 - 2G_N \mathcal{M}/r} \left( 2 \mathcal{V} + 3 \mathcal{N} \right)\nonumber\\ &-& \frac{9}{r} \mathcal{N} - 3 (\rho+p) \rho^{\prime} \, ,  \label{eq:7c}
\end{eqnarray}
\end{subequations}
where a prime indicates derivative with respect to $r$, $A(r)=[1-2G_{N}\mathcal{M}(r)/r]^{-1}$ and the effective energy density and pressure, respectively, are given as:
\begin{subequations}
\label{eq:3}
\begin{eqnarray}
\rho_{eff}  &=& \rho \left( 1 + \frac{\rho}{2\lambda} \right) + \frac{\mathcal{V}}{\lambda}  \, , \label{eq:3a} \\
p_{eff} &=& p \left(1 + \frac{\rho}{\lambda} \right) + \frac{\rho^{2}}{2\lambda} + \frac{\mathcal{V}}{3\lambda} + \frac{\mathcal{N}}{\lambda} \, . \label{eq:3b}
\end{eqnarray}
\end{subequations}
Even though we will not consider exterior solutions, we must anyway take into account the information provided by the Israel-Darmois (ID) matching condition, which for the case under study can be written as\cite{gm}:
\begin{equation}
  \label{eq:9}
  (3/2) \rho^2(R) + \mathcal{V}^-(R) + 3 \mathcal{N}^-(R) = \mathcal{V}^+(R) + 3 \mathcal{N}^+(R) \, ,
\end{equation}
where the superscript $-(+)$ denotes the interior (exterior) values of the different quantities at the surface of the star, and we also assumed that $\rho(r> R) =0$.

A desirable property we want in our solutions is a \emph{Schwarzschild
  exterior}, which can be easily accomplished under the boundary
conditions $\mathcal{V}^+(R) = 0 =\mathcal{N}^+(R)$, as
for them the simplest solution that arises from Eq.~\eqref{eq:7c} is
the trivial one: $\mathcal{V}(r \geq R) = 0 =\mathcal{N}(r
\geq R)$. Thus, for the purposes of this paper, we will refer
hereafter to the restricted ID matching condition given by:
\begin{equation}
  \label{eq:28}
    (3/2) \rho^2(R) + \mathcal{V}^-(R) + 3 \mathcal{N}^-(R) = 0 \, .
\end{equation}

For completeness, we just note that the exterior solutions of the metric functions are given by the well known expressions $B(r) = A^{-1}(r) = 1 - 2G_N M/r$.

Finally, an important point to remark is that \emph{the only interior solution of the nonlocal anisotropic stress under the conditions of a Schwarzschild exterior, and non-constant density with $\rho(R) = 0$, which are the conditions we expect to have in realistic stars, is the trivial one: $\mathcal{N}(r) \equiv 0$} (see\cite{Garcia-Aspeitia:2014pna} for details). Implying that Eq. \eqref{eq:28} can be written as:
\begin{equation}
  \label{eq:29}
    -(3/2) \rho^2(R) = \mathcal{V}^-(R) \, ,
\end{equation}
with the aim of maintain a Schwarzschild exterior.

\section{The modified Lane-Emden equation} \label{LES}

In principle, we should just numerically evolve Eqs. \eqref{eq:7}, but as we have to deal with weak gravity we find more appropriate to evolve the weak field limit of such system of equations, which by the way provides important technical simplifications that let us to have more physical insight. In order to get a star as real as possible and find the LE equation in the case of brane stars, we start imposing the following \emph{minimal conditions}:

\begin{enumerate}[(a)]

\item The radius $R$ is fixed, with $\rho(r)=0$ for $r>R$\cite{Garcia-Aspeitia:2014pna,Linares:2015fsa}.

\item The pressure vanishes at the surface and in the exterior of the star, and the $p(r)=0$ for $r\geq R$\cite{Garcia-Aspeitia:2014pna,Linares:2015fsa}.

\item The star is described by the polytropic equation $p=K\rho^{(1+n)/n}$, where $n$ is the polytropic index with $n\geq0$.

\item The pressure is negligible compared with the energy density $p\ll\rho$.

\item We assume the following relation $4\pi r^{3}p_{eff}\ll\mathcal{M}$, between effective variables.

\item The gravitational potential in terms of the effective mass is negligible, $2G_{N}\mathcal{M}/r\ll1$.

\end{enumerate}

Conditions (a) and (b) are the conventional wisdom, being physically reasonable assumptions for stellar configurations; both conditions are not imposed in the dynamical equations, however we expect that are satisfied in order to obtain a real star. For instance, in the case of condition (c) we propose a polytropic equation, which is the most similar component to a real star, condition (d) is necessary for the Newtonian approach, conditions (e) and (f) are similarly necessary for the Newtonian approach but is also important to make the comparison between the terms which generate the effective pressure and effective mass.

To begin with, we observe that under conditions (c)-(f) and from Eq. \eqref{eq:7b} we have:
\begin{equation}
r^{2}p^{\prime}=-G_{N}\mathcal{M}\rho,
\end{equation}
differentiating we found
\begin{equation}
\frac{d}{dr}\left(\frac{r^{2}}{\rho}\frac{dp}{dr}\right)=-4\pi G_{N}\rho_{eff}. \label{eqdiff9}
\end{equation}
Considering the following change of variables\cite{chandrasekhar1958introduction,*weinberg1972gravitation}:
\begin{subequations}
  \label{eq:20}
\begin{eqnarray}
  r =\left(\frac{K(n+1)}{4\pi G_{N}}\right)^{1/2}\rho(0)^{(1-n)/2n}\zeta,  \\
  \rho=\rho(0)\theta^{n}, \; p=K\rho(0)^{(n+1)/n}\theta^{n+1}, \label{neweq12}
\end{eqnarray}
\end{subequations}
and substituting in Eq. \eqref{eqdiff9} it is possible to write the LE equation modified by the presence of branes
\begin{equation}
\frac{1}{\zeta^{2}}\frac{d}{d\zeta}\zeta^{2}\frac{d\theta}{d\zeta}+\theta^{n}+\bar{\rho}(\theta^{2n}+\bar{\mathcal{V}}(\theta)_{n})=0, \label{LE}
\end{equation}
where $\bar{\mathcal{V}}(\theta)_{n}\equiv2\mathcal{V}(\theta)_{n}/\rho(0)^{2}$, $\bar{\rho}\equiv\rho(0)/2\lambda$. In addition from Eq. \eqref{eq:7d} and the renaming of variables \eqref{eq:20}, we obtain the dimensionless mass equation as:
\begin{equation}
\frac{d\bar{\mathcal{M}}}{d\zeta}-\theta^{n}-\bar{\rho}(\theta^{2n}+\bar{\mathcal{V}}(\theta)_{n})=0, \label{mass}
\end{equation}
or in quadratures
\begin{equation}
\bar{\mathcal{M}}=\int_{0}^{\xi_R}\theta^{n}d\zeta+\bar{\rho}\int_{0}^{\xi_R}(\theta^{2n}+\bar{\mathcal{V}}(\theta)_{n})d\zeta, \label{mass}
\end{equation}
where $\bar{\mathcal{M}}\equiv G_{N}^{3/2}\rho(0)^{-(3-n)/2n}\mathcal{M}/((4\pi)^{1/3}K(n+1))^{3/2}$. It is straightforward to see that the non-brane limit is recovered when $\bar{\rho}\to0$, in Eqs. \eqref{LE}-\eqref{mass}. The opposite case is the brane domination terms limit, when $\bar{\rho}\gg1$, obtaining the following equations of motion:
\begin{subequations}
\begin{eqnarray}
\frac{1}{\zeta^{2}}\frac{d}{d\zeta}\zeta^{2}\frac{d\theta}{d\zeta}+\bar{\rho}[\theta^{2n}+\bar{\mathcal{V}}(\theta)_{n}]=0,  \label{LEH} \\
\frac{d\bar{\mathcal{M}}}{d\zeta}-\bar{\rho}[\theta^{2n}+\bar{\mathcal{V}}(\theta)_{n}]=0.
\end{eqnarray}
\end{subequations}
Now, it is necessary to find the explicit functional form of the nonlocal energy density from the conservation equation \eqref{eq:7c} and with the help of conditions (d)-(f), then we have:
\begin{equation}
r^{2}\mathcal{V}^{\prime}+4G_{N}\mathcal{M}\mathcal{V}=-3(\rho+p)r^{2}\rho^{\prime},
\end{equation}
differentiating and rearranging with the use of Eqs. \eqref{eq:20} we obtain in general the following first order differential equation:
\begin{eqnarray}
\frac{d\bar{\mathcal{V}}_{n}}{d\theta}-\chi_{n}\bar{\mathcal{V}_{n}}=-6n\left[\theta^{2n-1}+\frac{\chi_{n}}{4(n+1)}\theta^{2n}\right],
\end{eqnarray}
whose solution can be computed through the following integral:
\begin{eqnarray}
\bar{\mathcal{V}}_n&=&-6n\exp(\chi_n\theta)\int\left[\theta^{2n-1}+\frac{\chi_{n}}{4(n+1)}\theta^{2n}\right]\exp(-\chi_n\theta)d\theta\nonumber\\&&+\mathcal{C}\exp(\chi_n\theta),
\end{eqnarray}
where we have also introduced the dimensionless quantities, $\chi_{n}\equiv4(n+1)K\rho(0)^{1/n}$ for $n\neq0$, $\mathcal{C}$ s an integration constant and $\bar{\mathcal{V}}_{n}$ is function of $\theta$, which in turn is a function of $r$. The solutions of the previous differential equation, without loss of generality, can be written as:
\begin{eqnarray}
\bar{\mathcal{V}}(\theta)_{n}&=&\frac{6n}{\chi^{2n}_{n}}\exp(\chi_{n}\theta)\left\lbrace\Gamma(2n,\chi_{n}\theta)+\frac{\Gamma(2n+1,\chi_{n}\theta)}{4(n+1)}\right\rbrace\nonumber\\&+&\mathcal{C}_{1}\exp(\chi_{n}\theta), \label{V1}
\end{eqnarray}
for $n\geq1/2$, being $\Gamma(x,y)$ the incomplete gamma function. Also we have:
\begin{eqnarray}
\bar{\mathcal{V}}(\theta)_{n}&=&6n\exp(\chi_{n}\theta)\left\lbrace \frac{\Gamma(2(1-n),\chi_{n}\theta)}{\chi_{n}^{2(1-n)}}+\frac{\Gamma(2n+1,\chi_{n}\theta)}{4(n+1)\chi_{n}^{2n}}\right\rbrace\nonumber\\&+&\mathcal{C}_{2}\exp(\chi_{n}\theta), \label{V2}
\end{eqnarray}
for $0< n<1/2$. In both cases, $\mathcal{C}_{1}$ and $\mathcal{C}_{2}$ are integration constants associated with the initial condition. Notice that in principle, the modified LE equation do not accept solutions for $n=0$, due to the divergence of the nonlocal energy density $\bar{\mathcal{V}}_{n}$; this would imply and unstable and non compact stellar configuration, being a prediction of this model.

Particularly, \emph{low energy} stars like dwarf stars can be modeled in this context and now we are in position to determine how the brane effects, provide extra dynamics in the interior of a star. It is important to mention that white dwarfs can be modeled by the polytropic index $n=3$, and neutron stars by polytropes with an index in the range $n=0.5-1$. However, in the case of neutron stars weak-field approximation is not sufficient to make a general description of these stars; it is necessary add the corrections provided by GR with the full modified TOV equation.

\section{Numerical Solutions for dwarf stars} \label{NA}

Let us start studying a dwarf star using the modified LE equation; we observe from Eq. \eqref{LE} that the free parameters are $\bar{\rho}$ and $\chi_{n}$, related with the central energy density of the star, brane tension and with the polytropic constant.

Our analysis show that central energy density and the polytropic constant are redundant, because in particular depends of the characteristics of each star, then, we fix by hand, the value of $\chi_{n}$, where are encoded both parameters. In this case, we boarded the region $\chi_{3}=10$, due that orders of magnitude greater cause divergences which implies non-compact configurations, this results in the following dependence: $K=5 \rho(0)^{-1/3}/8$ . Therefore, we only explore the limit case, when the minimal requirements (a)-(f) are fulfilled.

\subsection{Physical initial conditions}

Some physical initial conditions for the star are important; for this reason we start showing the equations of kinetic energy density an pressure of electrons of the dwarf star as\cite{chandrasekhar1958introduction,*weinberg1972gravitation}:
\begin{eqnarray}
 e&=&\frac{8\pi}{(2\pi)^3}\int_{0}^{k_{F}}[(k^2+m_{e}^{2})^{1/2}-m_{e}]k^{2}dk, \label{other}\\
 p&=&\frac{8\pi}{3(2\pi)^3}\int_{0}^{k_{F}}\frac{k^2}{(k^2+m_{e}^2)^{1/2}}k^2dk, \label{other1}
\end{eqnarray}
where the momenta is between $k$ and $k+dk$, being $k_{F}$ the maximum momentum and $m_{e}$ is related with the electron mass. From Eqs. \eqref{other} and \eqref{other1}, we obtain for dwarf stars with index $n=3$, the following conditions:
\begin{subequations}
\begin{eqnarray}
 e&=&3p, \;\; p=\frac{1}{12\pi^{2}}\left(\frac{3\pi^{2}\rho}{m_{N}\mu}\right)^{4/3}, \\
 K&=&\frac{1}{12\pi^{2}}\left(\frac{3\pi^{2}}{m_{N}\mu}\right)^{4/3}, \label{neweq1}
\end{eqnarray}
\end{subequations}
where $\mu$ is the number of nucleons per electron and $m_{N}$ is the nucleon mass\cite{chandrasekhar1958introduction,*weinberg1972gravitation}.

\subsection{Results of the numerical solutions} \label{RNS}

\begin{figure}[htbp]
\centering
\begin{tabular}{cc}
\includegraphics[scale=0.4]{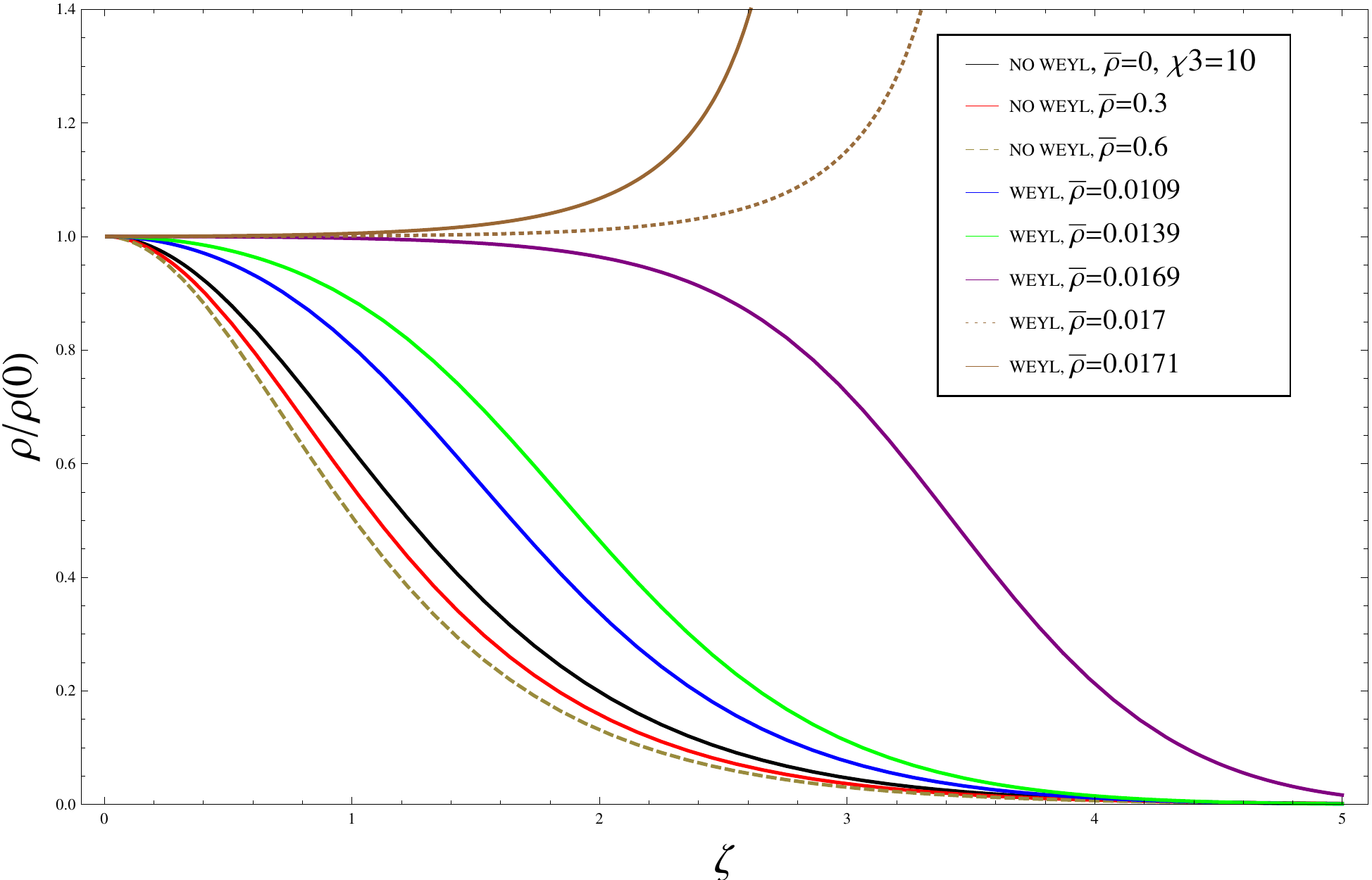}   \\
\includegraphics[scale=0.43]{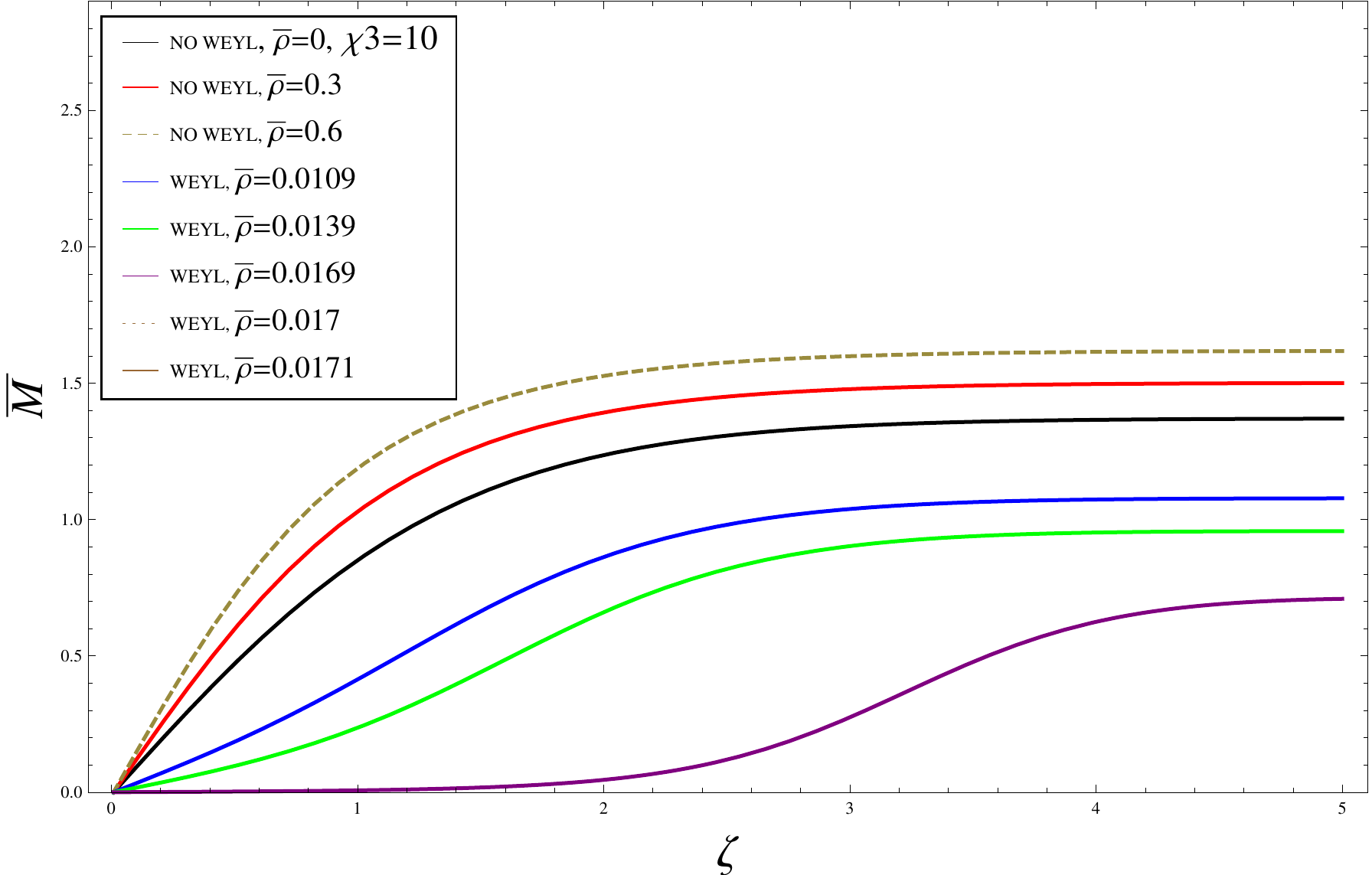}  
\end{tabular}
\caption{Numerical solution of Eqs. \eqref{LE}-\eqref{mass} for white dwarfs with polytropic index $n=3$. Here we show the energy density (Top) and effective mass (Bottom) of the white dwarf stars. Notice the sensitivity to the term $\bar{\rho}$, when we \emph{turn on} the Weyl terms. When Weyl terms are dominant, the stellar configuration is unstable, causing that conditions (a) and (b) are not fulfilled. See the text for more details.} 
\label{fig1}
\end{figure}

To begin with, we show the numerical solutions implemented for dwarf stars showing the behavior of energy density and mass profiles in Fig. \ref{fig1} (top and bottom). We implement the usual initial conditions: $\theta(0)=1$, $d\theta(0)/d\zeta=0$ and $\bar{\mathcal{M}}(0)=0$, for $n=3$, as in textbook case\cite{chandrasekhar1958introduction,*weinberg1972gravitation} and $\bar{\mathcal{V}}(0)=0$ considering an inward integration.

\begin{figure}[htbp]
\centering
\begin{tabular}{cc}
\includegraphics[scale=0.4]{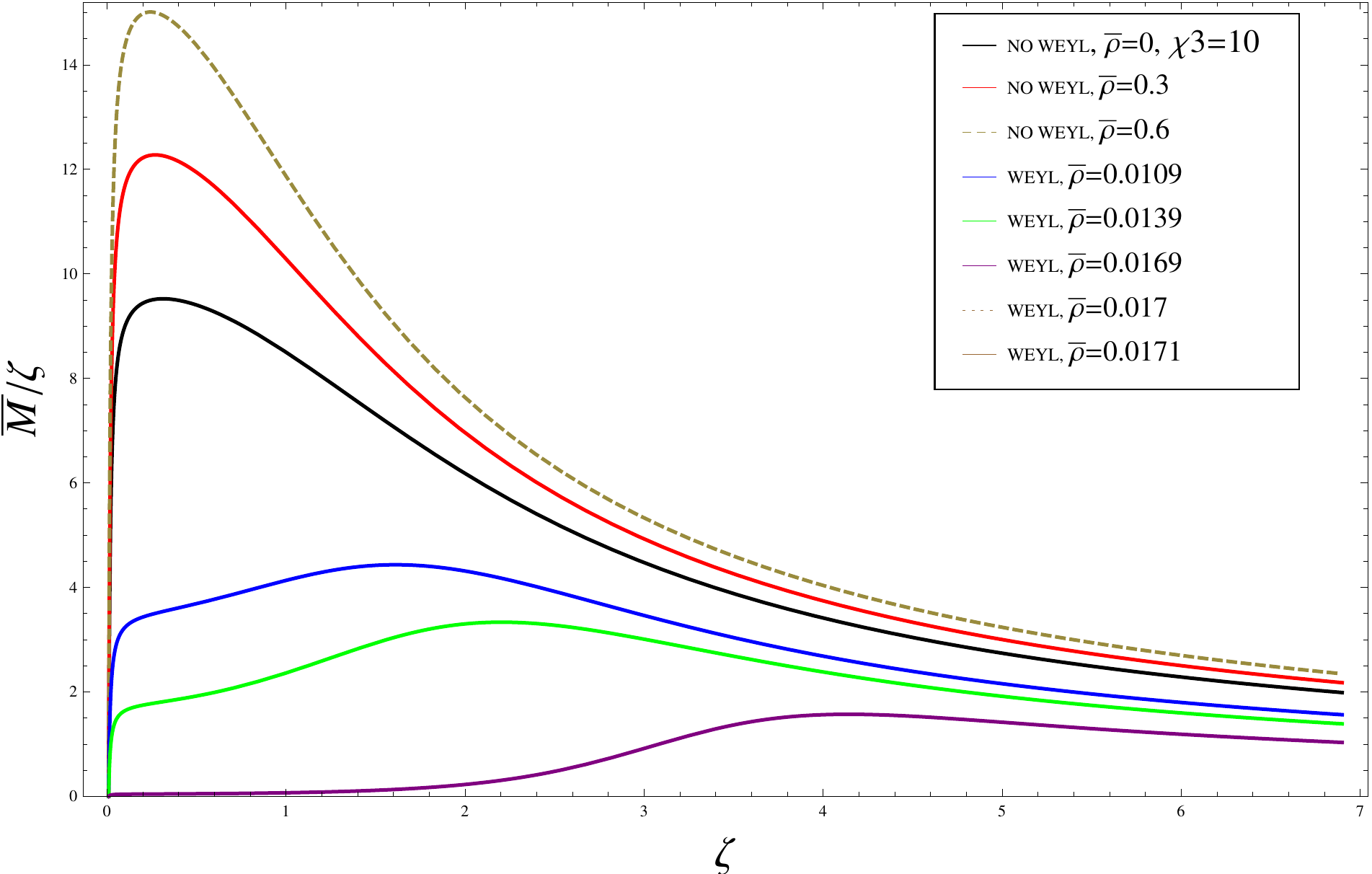}
\end{tabular}
\caption{Numerical solution of Eqs. \eqref{LE}-\eqref{mass} for white dwarfs with polytropic index $n=3$. Here we show the compactness $\bar{\mathcal{M}}/\zeta$ with the presence of Weyl terms and without. The presence of both terms generates a less compactness stellar configuration with a maximum displaced in comparison with the other cases. Notice how $\bar{\rho}=0.0169$ is the constriction for the plots. See the text for more details.} 
\label{fig2}
\end{figure}

We start showing the non-brane case as a benchmark, adding first, only the quadratic part of the energy-momentum tensor. Under this assumption, we predict a lower energy density compared with the non-brane case (see Fig. \ref{fig1}, top). Clearly, the stellar configuration is more massive for a similar radius to the previous case (see Fig. \ref{fig1}, bottom). Also, we present the compactness plot (see Fig. \ref{fig2}), which show the different behaviors with different values of brane terms. It is notorious how we have a most compactness configuration when the presence of quadratic terms predicted by branes plays an important role. Clearly, this is an incomplete analysis due to lack of Weyl terms, however in the following, we took on the task of presenting the nonlocal terms.

When we \emph{turn on} the Weyl terms, these cause higher energy densities and smaller masses in comparison with the case of non-branes, while we increase the presence of extra terms, the effects are accentuated, causing a non-compact configuration, \emph{i.e.}, conditions (a) and (b) are not fulfilled. In this sense, it is possible to note that $\bar{\rho}=0.016$, is the higher bound to have a stable stellar configuration; noticing that when we exceed this bound we have an unstable star, implying a non real stellar structure (see Fig. \ref{fig1} top and bottom and Fig. \ref{fig2}).

Under considerations of a stable star configuration which meets the minimal condition (a)-(g), it is possible to find the brane tension bound as $\lambda\gtrsim29.585 \; \rho(0)$, where $\rho(0)$ corresponds to the central energy density of the dwarf star. With the aim of compare with astrophysical data, in Table \ref{tab:wdwarfs} we show ten dwarf stars collected by the catalogues reported in Refs.\cite{wd1,*40erib,*sirius}, mainly emphasizing the values of mass, radius and central density; then for the samples of Table \ref{tab:wdwarfs} and under the assumption that observed white dwarfs must belong to a family of equilibrium configuration without an anomalous behavior, we have that the average value of brane tension must be: $\langle\lambda\rangle\gtrsim84.818\;\rm MeV^4$, with a standard deviation $\sigma\simeq82.021\;\rm MeV^4$, showing too much dispersion in the set of the sample. This is attributed to the marked differences between dwarf stars. In addition, notice how our results are weaker than other astrophysical data\cite{gm,Garcia-Aspeitia:2014pna,Linares:2015fsa}, however, it remains above the levels provided by cosmological bounds like nucleosynthesis\cite{Holanda:2013doa,*Barrow:2001pi,*Brax:2003fv}.

\begin{table}[htp]
\caption{\label{tab:wdwarfs} From left to right the columns read; name of the star, mass in solar units $M_{\odot}$, radius in $R_{\odot}$, density as $\rho(0) = 3M/4\pi
  R^3$ in ${\rm MeV}^4$ and brane tension in ${\rm MeV}^4$ deduced from the constraint mentioned above; using a catalogue of several white dwarfs reported in\cite{wd1,*40erib,*sirius}. See the text for more details.}
\begin{ruledtabular}
\begin{tabular}{c c c c c c }
White Dwarf & Mass ($M_{\odot}$) & Radius ($R_{\odot}$) & $
\rho(0)$ (${\rm MeV}^4$) & $
\lambda$ (${\rm MeV}^4$) \\
Sirius B & 1.034 & 0.0084 & 10.5993 & 313.588 \\
Procyon B & 0.604 & 0.0096 & 4.1478 & 122.715 \\
40 Eri B & 0.501 & 0.0136 & 1.21009 & 35.801\\
EG 50 & 0.50 & 0.0104 & 2.70063 &  79.900\\
GD 140 & 0.79 & 0.0085 & 7.81565 & 231.232\\
CD-38 10980 & 0.74 & 0.01245 & 2.3298 & 68.928\\
W485A & 0.59 & 0.0150 & 1.06212 & 31.423 \\
G154-B5B & 0.46 &0.0129 & 1.3006 & 38.4793 \\
LP 347-6 & 0.56 & 0.0124 & 1.7827 & 52.7426 \\
G181-B5B & 0.54 & 0.0125 & 1.6781 & 49.6479 \\
WD1550+130 & 0.535 & 0.0211 & 0.3456 & 10.2266 \\
Stein \; 2051B &0.48 & 0.0111 & 2.13023 & 63.0229 \\
G107-70AB &0.65 & 0.0127 & 1.926 & 56.9807 \\
L268-92 & 0.70 & 0.0149 & 1.28438 & 37.9984 \\
G156-64 &0.59 & 0.0110 & 2.69047 & 79.5976
\end{tabular}
\end{ruledtabular}
\end{table}

\section{Conclusions and Remarks} \label{C}

The presented analysis of weak field maintaining the brane terms, conducted by the LE equation, show us the new behavior of density, mass and compactness for stars with polytropic matter as content. The research developed in this paper, show how dwarf stars are sensitive to the Weyl terms causing a non-compact configuration under particular conditions, implying a non real star. It should be mentioned that the only existence of quadratic terms to the energy momentum tensor show a less dense and more massive star compared to the non-brane case. In general, when we \emph{turn on} also the Weyl contributions, the star rather suggest a behavior of higher energy density and lower mass, beyond standard GR, which is discussed in \ref{RNS}. Significantly, there is a physical limit to the parameters $\bar{\rho}$ and $\chi_{n}$ (see Fig. \ref{fig1} and \ref{fig2}) such that meets the minimum requirements for a stable star, where in this case must be: $\lambda\gtrsim29.585 \; \rho(0)$ and $K=5/(8\rho(0)^{1/3})$. Taking astrophysical data of a sample of white dwarfs, it is possible to establish an average bound of brane tension as shown in the previous section: $\langle\lambda\rangle\gtrsim84.818\;\rm MeV^4$, with a standard deviation $\sigma\simeq82.021\;\rm MeV^4$ and the average of the polytropic constant must be constrained as $\langle K\rangle\simeq0.508 \; \rm MeV^{-4/3}$ with a standard deviation $\sigma\simeq0.142;\rm MeV^{-4/3}$. It is important to remark how the previous values are necessary to fulfill the minimal requirements to obtain a stable star; \emph{i.e.} a real stellar configuration.

It is important to clarify that nonlocal terms caused by Weyl terms  are gravitons that escape to the fifth dimension, causing that star does not have a compact configuration as they begin to dominate. Excess of Weyl terms cause that does not satisfies the conditions (a) and (b), as we show along the paper. In fact, the Weyl terms eventually generate a divergence for a given radius. Although, we can use this disadvantage for us to quantify the minimum value required for brane tension which is shown in our conclusions.

In addition, it is worth mentioning that modifications to the LE equation \emph{prohibit} the case $n=0$ (at least for the case where $\mathcal{C}_{1}\neq0$ or $\mathcal{C}_{2}\neq0$), for a stable stellar configuration unlike that predicted by the non-branes limit. This is caused due to divergence in the $\chi_{n}$ term in the central energy density; causing that conditions (a) and (b) are not fulfilled.

Despite we are treating a weak gravitational limit and the brane effects aren't accentuated strongly in the dynamic; it is possible to extract relevant information about the constraint of brane tension, establishing an exclusion limit of the theory, taking as a premise the stability of the dwarf star. Strong evidence of branes can be found on the direct observation of the compactness of a dwarf star, when are compared the predictions of GR and Branes theory, bearing in mind the technical challenges of this endeavor, due to the subtle brane effects.

Finally we suggest that studies of neutron stars can give us better constraints and even evidence of the existence of extra dynamic which come from brane theories. In the case of neutron stars, part of the machinery has been studied in Ref.\cite{Garcia-Aspeitia:2014pna,Linares:2015fsa}, still as the most general way to treat this type of stars in a strong gravitational field. However this is a work that will be presented elsewhere.

\begin{acknowledgements}
The author acknowledge the suggestions of the anonymous referee and the enlightening conversation with F. Linares. Also the author is grateful for the support provided by SNI-M\'exico, CONACyT research fellow and  Instituto Avanzado de Cosmolog\'ia (IAC) collaborations.
\end{acknowledgements}

\bibliography{librero1}

\begin{thebibliography}{38}%
\makeatletter
\providecommand \@ifxundefined [1]{%
 \@ifx{#1\undefined}
}%
\providecommand \@ifnum [1]{%
 \ifnum #1\expandafter \@firstoftwo
 \else \expandafter \@secondoftwo
 \fi
}%
\providecommand \@ifx [1]{%
 \ifx #1\expandafter \@firstoftwo
 \else \expandafter \@secondoftwo
 \fi
}%
\providecommand \natexlab [1]{#1}%
\providecommand \enquote  [1]{``#1''}%
\providecommand \bibnamefont  [1]{#1}%
\providecommand \bibfnamefont [1]{#1}%
\providecommand \citenamefont [1]{#1}%
\providecommand \href@noop [0]{\@secondoftwo}%
\providecommand \href [0]{\begingroup \@sanitize@url \@href}%
\providecommand \@href[1]{\@@startlink{#1}\@@href}%
\providecommand \@@href[1]{\endgroup#1\@@endlink}%
\providecommand \@sanitize@url [0]{\catcode `\\12\catcode `\$12\catcode
  `\&12\catcode `\#12\catcode `\^12\catcode `\_12\catcode `\%12\relax}%
\providecommand \@@startlink[1]{}%
\providecommand \@@endlink[0]{}%
\providecommand \url  [0]{\begingroup\@sanitize@url \@url }%
\providecommand \@url [1]{\endgroup\@href {#1}{\urlprefix }}%
\providecommand \urlprefix  [0]{URL }%
\providecommand \Eprint [0]{\href }%
\providecommand \doibase [0]{http://dx.doi.org/}%
\providecommand \selectlanguage [0]{\@gobble}%
\providecommand \bibinfo  [0]{\@secondoftwo}%
\providecommand \bibfield  [0]{\@secondoftwo}%
\providecommand \translation [1]{[#1]}%
\providecommand \BibitemOpen [0]{}%
\providecommand \bibitemStop [0]{}%
\providecommand \bibitemNoStop [0]{.\EOS\space}%
\providecommand \EOS [0]{\spacefactor3000\relax}%
\providecommand \BibitemShut  [1]{\csname bibitem#1\endcsname}%
\let\auto@bib@innerbib\@empty
\bibitem [{\citenamefont {Chandrasekhar}(1985)}]{Chandrasekhar:1985kt}%
  \BibitemOpen
  \bibfield  {author} {\bibinfo {author} {\bibfnamefont {S.}~\bibnamefont
  {Chandrasekhar}},\ }\href@noop {} {\emph {\bibinfo {title} {{The mathematical
  theory of black holes}}}}\ (\bibinfo  {publisher} {Oxford University Press},\
  \bibinfo {year} {1985})\BibitemShut {NoStop}%
\bibitem [{\citenamefont {Tolman}(1987)}]{tolman1987relativity}%
  \BibitemOpen
  \bibfield  {author} {\bibinfo {author} {\bibfnamefont {R.~C.}\ \bibnamefont
  {Tolman}},\ }\href@noop {} {\emph {\bibinfo {title} {Relativity,
  thermodynamics, and cosmology}}}\ (\bibinfo  {publisher} {Dover Publications.
  com},\ \bibinfo {year} {1987})\BibitemShut {NoStop}%
\bibitem [{\citenamefont {Oppenheimer}\ and\ \citenamefont
  {Volkoff}(1939)}]{oppenheimer1939massive}%
  \BibitemOpen
  \bibfield  {author} {\bibinfo {author} {\bibfnamefont {J.~R.}\ \bibnamefont
  {Oppenheimer}}\ and\ \bibinfo {author} {\bibfnamefont {G.~M.}\ \bibnamefont
  {Volkoff}},\ }\href@noop {} {\bibfield  {journal} {\bibinfo  {journal}
  {Physical Review}\ }\textbf {\bibinfo {volume} {55}},\ \bibinfo {pages} {374}
  (\bibinfo {year} {1939})}\BibitemShut {NoStop}%
\bibitem [{\citenamefont
  {Chandrasekhar}(1958)}]{chandrasekhar1958introduction}%
  \BibitemOpen
  \bibfield  {author} {\bibinfo {author} {\bibfnamefont {S.}~\bibnamefont
  {Chandrasekhar}},\ }\href@noop {} {\emph {\bibinfo {title} {An introduction
  to the study of stellar structure}}},\ Vol.~\bibinfo {volume} {2}\ (\bibinfo
  {publisher} {DoverPublications. com},\ \bibinfo {year} {1958})\BibitemShut
  {NoStop}%
\bibitem [{\citenamefont {Weinberg}(1972)}]{weinberg1972gravitation}%
  \BibitemOpen
  \bibfield  {author} {\bibinfo {author} {\bibfnamefont {S.}~\bibnamefont
  {Weinberg}},\ }\href@noop {} {\emph {\bibinfo {title} {Gravitation and
  cosmology: principles and applications of the general theory of
  relativity}}},\ Vol.~\bibinfo {volume} {1}\ (\bibinfo  {publisher} {Wiley New
  York},\ \bibinfo {year} {1972})\BibitemShut {NoStop}%
\bibitem [{\citenamefont {Germani}\ and\ \citenamefont {Maartens}(2001)}]{gm}%
  \BibitemOpen
  \bibfield  {author} {\bibinfo {author} {\bibfnamefont {C.}~\bibnamefont
  {Germani}}\ and\ \bibinfo {author} {\bibfnamefont {R.}~\bibnamefont
  {Maartens}},\ }\href@noop {} {\bibfield  {journal} {\bibinfo  {journal}
  {Physical Review D}\ }\textbf {\bibinfo {volume} {64}},\ \bibinfo {pages}
  {124010} (\bibinfo {year} {2001})}\BibitemShut {NoStop}%
\bibitem [{\citenamefont {Ovalle}\ \emph {et~al.}(2013)\citenamefont {Ovalle},
  \citenamefont {Linares}, \citenamefont {Pasqua},\ and\ \citenamefont
  {Sotomayor}}]{Ovalle:2013vna}%
  \BibitemOpen
  \bibfield  {author} {\bibinfo {author} {\bibfnamefont {J.}~\bibnamefont
  {Ovalle}}, \bibinfo {author} {\bibfnamefont {F.}~\bibnamefont {Linares}},
  \bibinfo {author} {\bibfnamefont {A.}~\bibnamefont {Pasqua}}, \ and\ \bibinfo
  {author} {\bibfnamefont {A.}~\bibnamefont {Sotomayor}},\ }\href {\doibase
  10.1088/0264-9381/30/17/175019} {\bibfield  {journal} {\bibinfo  {journal}
  {Class. Quant. Grav.}\ }\textbf {\bibinfo {volume} {30}},\ \bibinfo {pages}
  {175019} (\bibinfo {year} {2013})},\ \Eprint {http://arxiv.org/abs/1304.5995}
  {arXiv:1304.5995 [gr-qc]} \BibitemShut {NoStop}%
\bibitem [{\citenamefont {Ovalle}\ and\ \citenamefont
  {Linares}(2013)}]{Ovalle:2013xla}%
  \BibitemOpen
  \bibfield  {author} {\bibinfo {author} {\bibfnamefont {J.}~\bibnamefont
  {Ovalle}}\ and\ \bibinfo {author} {\bibfnamefont {F.}~\bibnamefont
  {Linares}},\ }\href {\doibase 10.1103/PhysRevD.88.104026} {\bibfield
  {journal} {\bibinfo  {journal} {Phys. Rev.}\ }\textbf {\bibinfo {volume}
  {D88}},\ \bibinfo {pages} {104026} (\bibinfo {year} {2013})},\ \Eprint
  {http://arxiv.org/abs/1311.1844} {arXiv:1311.1844 [gr-qc]} \BibitemShut
  {NoStop}%
\bibitem [{\citenamefont {Ovalle}\ \emph {et~al.}(2014)\citenamefont {Ovalle},
  \citenamefont {Gergely},\ and\ \citenamefont {Casadio}}]{Ovalle:2014uwa}%
  \BibitemOpen
  \bibfield  {author} {\bibinfo {author} {\bibfnamefont {J.}~\bibnamefont
  {Ovalle}}, \bibinfo {author} {\bibfnamefont {L.~Á.}\ \bibnamefont {Gergely}},
  \ and\ \bibinfo {author} {\bibfnamefont {R.}~\bibnamefont {Casadio}},\
  }\href@noop {} {\  (\bibinfo {year} {2014})},\ \Eprint
  {http://arxiv.org/abs/1405.0252} {arXiv:1405.0252 [gr-qc]} \BibitemShut
  {NoStop}%
\bibitem [{\citenamefont {Garcia-Aspeitia}\ \emph {et~al.}(2014)\citenamefont
  {Garcia-Aspeitia}, \citenamefont {Reyes-Ibarra}, \citenamefont {Ortiz},
  \citenamefont {Lopez-Dominguez},\ and\ \citenamefont
  {Hinojosa-Ruiz}}]{Garcia-Aspeitia:2014jda}%
  \BibitemOpen
  \bibfield  {author} {\bibinfo {author} {\bibfnamefont {M.~A.}\ \bibnamefont
  {Garcia-Aspeitia}}, \bibinfo {author} {\bibfnamefont {M.~J.}\ \bibnamefont
  {Reyes-Ibarra}}, \bibinfo {author} {\bibfnamefont {C.}~\bibnamefont {Ortiz}},
  \bibinfo {author} {\bibfnamefont {J.}~\bibnamefont {Lopez-Dominguez}}, \ and\
  \bibinfo {author} {\bibfnamefont {S.}~\bibnamefont {Hinojosa-Ruiz}},\
  }\href@noop {} {\  (\bibinfo {year} {2014})},\ \Eprint
  {http://arxiv.org/abs/1412.3496} {arXiv:1412.3496 [gr-qc]} \BibitemShut
  {NoStop}%
\bibitem [{\citenamefont {Garc\'ia-Aspeitia}\ and\ \citenamefont {Ure\~na
  Lopez}(2015)}]{Garcia-Aspeitia:2014pna}%
  \BibitemOpen
  \bibfield  {author} {\bibinfo {author} {\bibfnamefont {M.~A.}\ \bibnamefont
  {Garc\'ia-Aspeitia}}\ and\ \bibinfo {author} {\bibfnamefont {L.~A.}\
  \bibnamefont {Ure\~na Lopez}},\ }\href {\doibase
  10.1088/0264-9381/32/2/025014} {\bibfield  {journal} {\bibinfo  {journal}
  {Class. Quantum Grav.}\ }\textbf {\bibinfo {volume} {32}},\ \bibinfo {pages}
  {025014} (\bibinfo {year} {2015})},\ \Eprint
  {http://arxiv.org/abs/gr-qc:1405.3932} {arXiv:gr-qc:1405.3932} \BibitemShut
  {NoStop}%
\bibitem [{\citenamefont {Linares}\ \emph {et~al.}(2015)\citenamefont
  {Linares}, \citenamefont {Garc\'{i}a-Aspeitia},\ and\ \citenamefont {Ure\~na
  L\'opez}}]{Linares:2015fsa}%
  \BibitemOpen
  \bibfield  {author} {\bibinfo {author} {\bibfnamefont {F.~X.}\ \bibnamefont
  {Linares}}, \bibinfo {author} {\bibfnamefont {M.~A.}\ \bibnamefont
  {Garc\'{i}a-Aspeitia}}, \ and\ \bibinfo {author} {\bibfnamefont {L.~A.}\
  \bibnamefont {Ure\~na L\'opez}},\ }\href {\doibase
  10.1103/PhysRevD.92.024037} {\bibfield  {journal} {\bibinfo  {journal} {Phys.
  Rev. D}\ }\textbf {\bibinfo {volume} {92}},\ \bibinfo {pages} {024037}
  (\bibinfo {year} {2015})}\BibitemShut {NoStop}%
\bibitem [{\citenamefont {Perez-Lorenzana}(2005)}]{PerezLorenzana:2005iv}%
  \BibitemOpen
  \bibfield  {author} {\bibinfo {author} {\bibfnamefont {A.}~\bibnamefont
  {Perez-Lorenzana}},\ }\href {\doibase 10.1088/1742-6596/18/1/006} {\bibfield
  {journal} {\bibinfo  {journal} {J. Phys. Conf. Ser.}\ }\textbf {\bibinfo
  {volume} {18}},\ \bibinfo {pages} {224} (\bibinfo {year} {2005})},\ \Eprint
  {http://arxiv.org/abs/hep-ph/0503177} {arXiv:hep-ph/0503177 [hep-ph]}
  \BibitemShut {NoStop}%
\bibitem [{\citenamefont {Maartens}\ and\ \citenamefont {Koyama}(2010)}]{mk}%
  \BibitemOpen
  \bibfield  {author} {\bibinfo {author} {\bibfnamefont {R.}~\bibnamefont
  {Maartens}}\ and\ \bibinfo {author} {\bibfnamefont {K.}~\bibnamefont
  {Koyama}},\ }\href@noop {} {\bibfield  {journal} {\bibinfo  {journal} {Living
  Rev. Rel.}\ }\textbf {\bibinfo {volume} {13}},\ \bibinfo {pages} {5}
  (\bibinfo {year} {2010})},\ \Eprint {http://arxiv.org/abs/1004.3962}
  {arXiv:1004.3962 [hep-th]} \BibitemShut {NoStop}%
\bibitem [{\citenamefont {Randall}\ and\ \citenamefont
  {Sundrum}(1999{\natexlab{a}})}]{Randall-I}%
  \BibitemOpen
  \bibfield  {author} {\bibinfo {author} {\bibfnamefont {L.}~\bibnamefont
  {Randall}}\ and\ \bibinfo {author} {\bibfnamefont {R.}~\bibnamefont
  {Sundrum}},\ }\href {\doibase 10.1103/PhysRevLett.83.3370} {\bibfield
  {journal} {\bibinfo  {journal} {Phys. Rev. Lett.}\ }\textbf {\bibinfo
  {volume} {83}},\ \bibinfo {pages} {3370} (\bibinfo {year}
  {1999}{\natexlab{a}})},\ \Eprint {http://arxiv.org/abs/hep-ph/9905221}
  {arXiv:hep-ph/9905221} \BibitemShut {NoStop}%
\bibitem [{\citenamefont {Randall}\ and\ \citenamefont
  {Sundrum}(1999{\natexlab{b}})}]{Randall-II}%
  \BibitemOpen
  \bibfield  {author} {\bibinfo {author} {\bibfnamefont {L.}~\bibnamefont
  {Randall}}\ and\ \bibinfo {author} {\bibfnamefont {R.}~\bibnamefont
  {Sundrum}},\ }\href {\doibase 10.1103/PhysRevLett.83.4690} {\bibfield
  {journal} {\bibinfo  {journal} {Phys. Rev. Lett.}\ }\textbf {\bibinfo
  {volume} {83}},\ \bibinfo {pages} {4690} (\bibinfo {year}
  {1999}{\natexlab{b}})},\ \Eprint {http://arxiv.org/abs/hep-th/9906064}
  {arXiv:hep-th/9906064 [hep-th]} \BibitemShut {NoStop}%
\bibitem [{\citenamefont {McWilliams}(2010)}]{PhysRevLett.104.141601}%
  \BibitemOpen
  \bibfield  {author} {\bibinfo {author} {\bibfnamefont {S.~T.}\ \bibnamefont
  {McWilliams}},\ }\href {\doibase 10.1103/PhysRevLett.104.141601} {\bibfield
  {journal} {\bibinfo  {journal} {Phys. Rev. Lett.}\ }\textbf {\bibinfo
  {volume} {104}},\ \bibinfo {pages} {141601} (\bibinfo {year}
  {2010})}\BibitemShut {NoStop}%
\bibitem [{\citenamefont {Maartens}(2000)}]{Maartens:2000fg}%
  \BibitemOpen
  \bibfield  {author} {\bibinfo {author} {\bibfnamefont {R.}~\bibnamefont
  {Maartens}},\ }\href {\doibase 10.1103/PhysRevD.62.084023} {\bibfield
  {journal} {\bibinfo  {journal} {Phys. Rev.}\ }\textbf {\bibinfo {volume}
  {D62}},\ \bibinfo {pages} {084023} (\bibinfo {year} {2000})},\ \Eprint
  {http://arxiv.org/abs/hep-th/0004166} {arXiv:hep-th/0004166 [hep-th]}
  \BibitemShut {NoStop}%
\bibitem [{\citenamefont {Brax}\ \emph {et~al.}(2004)\citenamefont {Brax},
  \citenamefont {van~de Bruck},\ and\ \citenamefont {Davis}}]{Brax:2004xh}%
  \BibitemOpen
  \bibfield  {author} {\bibinfo {author} {\bibfnamefont {P.}~\bibnamefont
  {Brax}}, \bibinfo {author} {\bibfnamefont {C.}~\bibnamefont {van~de Bruck}},
  \ and\ \bibinfo {author} {\bibfnamefont {A.-C.}\ \bibnamefont {Davis}},\
  }\href {\doibase 10.1088/0034-4885/67/12/R02} {\bibfield  {journal} {\bibinfo
   {journal} {Rept. Prog. Phys.}\ }\textbf {\bibinfo {volume} {67}},\ \bibinfo
  {pages} {2183} (\bibinfo {year} {2004})},\ \Eprint
  {http://arxiv.org/abs/hep-th/0404011} {arXiv:hep-th/0404011 [hep-th]}
  \BibitemShut {NoStop}%
\bibitem [{\citenamefont {Brax}\ and\ \citenamefont {van~de
  Bruck}(2003)}]{Brax:2003fv}%
  \BibitemOpen
  \bibfield  {author} {\bibinfo {author} {\bibfnamefont {P.}~\bibnamefont
  {Brax}}\ and\ \bibinfo {author} {\bibfnamefont {C.}~\bibnamefont {van~de
  Bruck}},\ }\href {\doibase 10.1088/0264-9381/20/9/202} {\bibfield  {journal}
  {\bibinfo  {journal} {Class. Quant. Grav.}\ }\textbf {\bibinfo {volume}
  {20}},\ \bibinfo {pages} {R201} (\bibinfo {year} {2003})},\ \Eprint
  {http://arxiv.org/abs/hep-th/0303095} {arXiv:hep-th/0303095 [hep-th]}
  \BibitemShut {NoStop}%
\bibitem [{\citenamefont {Aspeitia}\ and\ \citenamefont
  {Matos}(2011)}]{Aspeitia:2009bj}%
  \BibitemOpen
  \bibfield  {author} {\bibinfo {author} {\bibfnamefont {M.~A.~G.}\
  \bibnamefont {Aspeitia}}\ and\ \bibinfo {author} {\bibfnamefont
  {T.}~\bibnamefont {Matos}},\ }\href {\doibase 10.1007/s10714-010-1093-2}
  {\bibfield  {journal} {\bibinfo  {journal} {Gen. Rel. Grav.}\ }\textbf
  {\bibinfo {volume} {43}},\ \bibinfo {pages} {315} (\bibinfo {year} {2011})},\
  \Eprint {http://arxiv.org/abs/0906.3278} {arXiv:0906.3278 [gr-qc]}
  \BibitemShut {NoStop}%
\bibitem [{\citenamefont {Garcia-Aspeitia}\ \emph {et~al.}(2012)\citenamefont
  {Garcia-Aspeitia}, \citenamefont {Magana},\ and\ \citenamefont
  {Matos}}]{GarciaAspeitia:2011xv}%
  \BibitemOpen
  \bibfield  {author} {\bibinfo {author} {\bibfnamefont {M.~A.}\ \bibnamefont
  {Garcia-Aspeitia}}, \bibinfo {author} {\bibfnamefont {J.}~\bibnamefont
  {Magana}}, \ and\ \bibinfo {author} {\bibfnamefont {T.}~\bibnamefont
  {Matos}},\ }\href {\doibase 10.1007/s10714-011-1294-3} {\bibfield  {journal}
  {\bibinfo  {journal} {Gen. Rel. Grav.}\ }\textbf {\bibinfo {volume} {44}},\
  \bibinfo {pages} {581} (\bibinfo {year} {2012})},\ \Eprint
  {http://arxiv.org/abs/1102.0825} {arXiv:1102.0825 [gr-qc]} \BibitemShut
  {NoStop}%
\bibitem [{\citenamefont {Castro}\ \emph {et~al.}(2014)\citenamefont {Castro},
  \citenamefont {Alloy},\ and\ \citenamefont {Menezes}}]{Castro:2014xza}%
  \BibitemOpen
  \bibfield  {author} {\bibinfo {author} {\bibfnamefont {L.~B.}\ \bibnamefont
  {Castro}}, \bibinfo {author} {\bibfnamefont {M.~D.}\ \bibnamefont {Alloy}}, \
  and\ \bibinfo {author} {\bibfnamefont {D.~P.}\ \bibnamefont {Menezes}},\
  }\href {\doibase 10.1088/1475-7516/2014/08/047} {\bibfield  {journal}
  {\bibinfo  {journal} {JCAP}\ }\textbf {\bibinfo {volume} {1408}},\ \bibinfo
  {pages} {047} (\bibinfo {year} {2014})},\ \Eprint
  {http://arxiv.org/abs/1403.1099} {arXiv:1403.1099 [nucl-th]} \BibitemShut
  {NoStop}%
\bibitem [{\citenamefont {Casadio}\ and\ \citenamefont
  {Ovalle}(2012)}]{Casadio:2012pu}%
  \BibitemOpen
  \bibfield  {author} {\bibinfo {author} {\bibfnamefont {R.}~\bibnamefont
  {Casadio}}\ and\ \bibinfo {author} {\bibfnamefont {J.}~\bibnamefont
  {Ovalle}},\ }\href {\doibase 10.1016/j.physletb.2012.07.041} {\bibfield
  {journal} {\bibinfo  {journal} {Phys. Lett.}\ }\textbf {\bibinfo {volume}
  {B715}},\ \bibinfo {pages} {251} (\bibinfo {year} {2012})},\ \Eprint
  {http://arxiv.org/abs/1201.6145} {arXiv:1201.6145 [gr-qc]} \BibitemShut
  {NoStop}%
\bibitem [{\citenamefont {Casadio}\ and\ \citenamefont
  {Ovalle}(2014)}]{Casadio:2012rf}%
  \BibitemOpen
  \bibfield  {author} {\bibinfo {author} {\bibfnamefont {R.}~\bibnamefont
  {Casadio}}\ and\ \bibinfo {author} {\bibfnamefont {J.}~\bibnamefont
  {Ovalle}},\ }\href {\doibase 10.1007/s10714-014-1669-3} {\bibfield  {journal}
  {\bibinfo  {journal} {Gen. Rel. Grav.}\ }\textbf {\bibinfo {volume} {46}},\
  \bibinfo {pages} {1669} (\bibinfo {year} {2014})},\ \Eprint
  {http://arxiv.org/abs/1212.0409} {arXiv:1212.0409 [gr-qc]} \BibitemShut
  {NoStop}%
\bibitem [{\citenamefont {Kapner}\ \emph {et~al.}(2007)\citenamefont {Kapner},
  \citenamefont {Cook}, \citenamefont {Adelberger}, \citenamefont {Gundlach},
  \citenamefont {Heckel} \emph {et~al.}}]{Kapner:2006si}%
  \BibitemOpen
  \bibfield  {author} {\bibinfo {author} {\bibfnamefont {D.}~\bibnamefont
  {Kapner}}, \bibinfo {author} {\bibfnamefont {T.}~\bibnamefont {Cook}},
  \bibinfo {author} {\bibfnamefont {E.}~\bibnamefont {Adelberger}}, \bibinfo
  {author} {\bibfnamefont {J.}~\bibnamefont {Gundlach}}, \bibinfo {author}
  {\bibfnamefont {B.~R.}\ \bibnamefont {Heckel}},  \emph {et~al.},\ }\href
  {\doibase 10.1103/PhysRevLett.98.021101} {\bibfield  {journal} {\bibinfo
  {journal} {Phys. Rev. Lett.}\ }\textbf {\bibinfo {volume} {98}},\ \bibinfo
  {pages} {021101} (\bibinfo {year} {2007})},\ \Eprint
  {http://arxiv.org/abs/hep-ph/0611184} {arXiv:hep-ph/0611184 [hep-ph]}
  \BibitemShut {NoStop}%
\bibitem [{\citenamefont {Alexeyev}\ \emph {et~al.}(2015)\citenamefont
  {Alexeyev}, \citenamefont {Rannu}, \citenamefont {Dyadina}, \citenamefont
  {Latosh},\ and\ \citenamefont {Turyshev}}]{Alexeyev:2015gja}%
  \BibitemOpen
  \bibfield  {author} {\bibinfo {author} {\bibfnamefont {S.}~\bibnamefont
  {Alexeyev}}, \bibinfo {author} {\bibfnamefont {K.}~\bibnamefont {Rannu}},
  \bibinfo {author} {\bibfnamefont {P.}~\bibnamefont {Dyadina}}, \bibinfo
  {author} {\bibfnamefont {B.}~\bibnamefont {Latosh}}, \ and\ \bibinfo {author}
  {\bibfnamefont {S.}~\bibnamefont {Turyshev}},\ }\href {\doibase
  10.7868/S0044451015060051} {\  (\bibinfo {year} {2015}),\
  10.7868/S0044451015060051},\ \Eprint {http://arxiv.org/abs/1501.04217}
  {arXiv:1501.04217 [gr-qc]} \BibitemShut {NoStop}%
\bibitem [{\citenamefont {Sakstein}\ \emph {et~al.}(2014)\citenamefont
  {Sakstein}, \citenamefont {Jain},\ and\ \citenamefont
  {Vikram}}]{Sakstein:2014nfa}%
  \BibitemOpen
  \bibfield  {author} {\bibinfo {author} {\bibfnamefont {J.}~\bibnamefont
  {Sakstein}}, \bibinfo {author} {\bibfnamefont {B.}~\bibnamefont {Jain}}, \
  and\ \bibinfo {author} {\bibfnamefont {V.}~\bibnamefont {Vikram}},\ }\href
  {\doibase 10.1142/S0218271814420024} {\bibfield  {journal} {\bibinfo
  {journal} {Int. J. Mod. Phys.}\ }\textbf {\bibinfo {volume} {D23}},\ \bibinfo
  {pages} {1442002} (\bibinfo {year} {2014})},\ \Eprint
  {http://arxiv.org/abs/1409.3708} {arXiv:1409.3708 [astro-ph.CO]} \BibitemShut
  {NoStop}%
\bibitem [{\citenamefont {García-Aspeitia}(2014)}]{GA2013}%
  \BibitemOpen
  \bibfield  {author} {\bibinfo {author} {\bibfnamefont {M.~A.}\ \bibnamefont
  {García-Aspeitia}},\ }\href@noop {} {\bibfield  {journal} {\bibinfo
  {journal} {Rev. Mex. Fis.}\ }\textbf {\bibinfo {volume} {60}},\ \bibinfo
  {pages} {205} (\bibinfo {year} {2014})},\ \Eprint
  {http://arxiv.org/abs/1306.1283} {arXiv:1306.1283 [gr-qc]} \BibitemShut
  {NoStop}%
\bibitem [{\citenamefont {Kudoh}\ \emph {et~al.}(2003)\citenamefont {Kudoh},
  \citenamefont {Tanaka},\ and\ \citenamefont {Nakamura}}]{Kudoh:2003xz}%
  \BibitemOpen
  \bibfield  {author} {\bibinfo {author} {\bibfnamefont {H.}~\bibnamefont
  {Kudoh}}, \bibinfo {author} {\bibfnamefont {T.}~\bibnamefont {Tanaka}}, \
  and\ \bibinfo {author} {\bibfnamefont {T.}~\bibnamefont {Nakamura}},\ }\href
  {\doibase 10.1103/PhysRevD.68.024035} {\bibfield  {journal} {\bibinfo
  {journal} {Phys. Rev.}\ }\textbf {\bibinfo {volume} {D68}},\ \bibinfo {pages}
  {024035} (\bibinfo {year} {2003})},\ \Eprint
  {http://arxiv.org/abs/gr-qc/0301089} {arXiv:gr-qc/0301089 [gr-qc]}
  \BibitemShut {NoStop}%
\bibitem [{\citenamefont {Cavaglia}(2003)}]{Cavaglia:2002si}%
  \BibitemOpen
  \bibfield  {author} {\bibinfo {author} {\bibfnamefont {M.}~\bibnamefont
  {Cavaglia}},\ }\href {\doibase 10.1142/S0217751X03013569} {\bibfield
  {journal} {\bibinfo  {journal} {Int. J. Mod. Phys.}\ }\textbf {\bibinfo
  {volume} {A18}},\ \bibinfo {pages} {1843} (\bibinfo {year} {2003})},\ \Eprint
  {http://arxiv.org/abs/hep-ph/0210296} {arXiv:hep-ph/0210296 [hep-ph]}
  \BibitemShut {NoStop}%
\bibitem [{\citenamefont {Holanda}\ \emph {et~al.}(2013)\citenamefont
  {Holanda}, \citenamefont {Silva},\ and\ \citenamefont
  {Dahia}}]{Holanda:2013doa}%
  \BibitemOpen
  \bibfield  {author} {\bibinfo {author} {\bibfnamefont {R.}~\bibnamefont
  {Holanda}}, \bibinfo {author} {\bibfnamefont {J.}~\bibnamefont {Silva}}, \
  and\ \bibinfo {author} {\bibfnamefont {F.}~\bibnamefont {Dahia}},\ }\href
  {\doibase 10.1088/0264-9381/30/20/205003} {\bibfield  {journal} {\bibinfo
  {journal} {Class. Quant. Grav.}\ }\textbf {\bibinfo {volume} {30}},\ \bibinfo
  {pages} {205003} (\bibinfo {year} {2013})},\ \Eprint
  {http://arxiv.org/abs/1304.4746} {arXiv:1304.4746 [astro-ph.CO]} \BibitemShut
  {NoStop}%
\bibitem [{\citenamefont {Barrow}\ and\ \citenamefont
  {Maartens}(2002)}]{Barrow:2001pi}%
  \BibitemOpen
  \bibfield  {author} {\bibinfo {author} {\bibfnamefont {J.~D.}\ \bibnamefont
  {Barrow}}\ and\ \bibinfo {author} {\bibfnamefont {R.}~\bibnamefont
  {Maartens}},\ }\href {\doibase 10.1016/S0370-2693(02)01552-6} {\bibfield
  {journal} {\bibinfo  {journal} {Phys. Lett.}\ }\textbf {\bibinfo {volume}
  {B532}},\ \bibinfo {pages} {153} (\bibinfo {year} {2002})},\ \Eprint
  {http://arxiv.org/abs/gr-qc/0108073} {arXiv:gr-qc/0108073 [gr-qc]}
  \BibitemShut {NoStop}%
\bibitem [{\citenamefont {Shiromizu}\ \emph {et~al.}(2000)\citenamefont
  {Shiromizu}, \citenamefont {Maeda},\ and\ \citenamefont {Sasaki}}]{sms}%
  \BibitemOpen
  \bibfield  {author} {\bibinfo {author} {\bibfnamefont {T.}~\bibnamefont
  {Shiromizu}}, \bibinfo {author} {\bibfnamefont {K.}~\bibnamefont {Maeda}}, \
  and\ \bibinfo {author} {\bibfnamefont {M.}~\bibnamefont {Sasaki}},\ }\href
  {http://link.aps.org/doi/10.1103/PhysRevD.62.024012} {\bibfield  {journal}
  {\bibinfo  {journal} {Phys. Rev. D}\ }\textbf {\bibinfo {volume} {62}},\
  \bibinfo {pages} {024012} (\bibinfo {year} {2000})}\BibitemShut {NoStop}%
\bibitem [{\citenamefont {Pavlidou}\ and\ \citenamefont
  {Tomaras}(2013)}]{Pavlidou:2013zha}%
  \BibitemOpen
  \bibfield  {author} {\bibinfo {author} {\bibfnamefont {V.}~\bibnamefont
  {Pavlidou}}\ and\ \bibinfo {author} {\bibfnamefont {T.~N.}\ \bibnamefont
  {Tomaras}},\ }\href@noop {} {\  (\bibinfo {year} {2013})},\ \Eprint
  {http://arxiv.org/abs/1310.1920} {arXiv:1310.1920 [astro-ph.CO]} \BibitemShut
  {NoStop}%
\bibitem [{\citenamefont {Balberg}\ and\ \citenamefont {Shapiro}(2000)}]{wd1}%
  \BibitemOpen
  \bibfield  {author} {\bibinfo {author} {\bibfnamefont {S.}~\bibnamefont
  {Balberg}}\ and\ \bibinfo {author} {\bibfnamefont {S.~L.}\ \bibnamefont
  {Shapiro}},\ }\href@noop {} {\bibfield  {journal} {\bibinfo  {journal} {arXiv
  preprint astro-ph/0004317}\ } (\bibinfo {year} {2000})}\BibitemShut {NoStop}%
\bibitem [{\citenamefont {Shipman}\ \emph {et~al.}(1997)\citenamefont
  {Shipman}, \citenamefont {Provencal}, \citenamefont {H{\o}g},\ and\
  \citenamefont {Thejll}}]{40erib}%
  \BibitemOpen
  \bibfield  {author} {\bibinfo {author} {\bibfnamefont {H.~L.}\ \bibnamefont
  {Shipman}}, \bibinfo {author} {\bibfnamefont {J.}~\bibnamefont {Provencal}},
  \bibinfo {author} {\bibfnamefont {E.}~\bibnamefont {H{\o}g}}, \ and\ \bibinfo
  {author} {\bibfnamefont {P.}~\bibnamefont {Thejll}},\ }\href@noop {}
  {\bibfield  {journal} {\bibinfo  {journal} {The Astrophysical Journal
  Letters}\ }\textbf {\bibinfo {volume} {488}},\ \bibinfo {pages} {L43}
  (\bibinfo {year} {1997})}\BibitemShut {NoStop}%
\bibitem [{\citenamefont {Holberg}\ \emph {et~al.}(1998)\citenamefont
  {Holberg}, \citenamefont {Barstow}, \citenamefont {Bruhweiler}, \citenamefont
  {Cruise},\ and\ \citenamefont {Penny}}]{sirius}%
  \BibitemOpen
  \bibfield  {author} {\bibinfo {author} {\bibfnamefont {J.~B.}\ \bibnamefont
  {Holberg}}, \bibinfo {author} {\bibfnamefont {M.}~\bibnamefont {Barstow}},
  \bibinfo {author} {\bibfnamefont {F.}~\bibnamefont {Bruhweiler}}, \bibinfo
  {author} {\bibfnamefont {A.}~\bibnamefont {Cruise}}, \ and\ \bibinfo {author}
  {\bibfnamefont {A.}~\bibnamefont {Penny}},\ }\href@noop {} {\bibfield
  {journal} {\bibinfo  {journal} {The Astrophysical Journal}\ }\textbf
  {\bibinfo {volume} {497}},\ \bibinfo {pages} {935} (\bibinfo {year}
  {1998})}\BibitemShut {NoStop}%
\end{thebibliography}%

\end{document}